\documentclass[aps,prl,twocolumn,tightenlines,superscriptaddress,showpacs,byrevtex]{revtex4}

\usepackage{graphicx}
\usepackage{dcolumn}
\usepackage{bm}

\graphicspath{{./}}

\def\Mbc {M_{\rm bc}}

\def\Dmb {\Delta M_{B}}

\begin{document}

\title{ 
 Observation of $B^{-} \to J/\psi \Lambda \bar{p}$ 
 and Searches for $B^{-} \to J/\psi \Sigma^{0} \bar{p}$ and  
 $B^{0} \to J/\psi p \bar{p}$ Decays }
\affiliation{Budker Institute of Nuclear Physics, Novosibirsk}
\affiliation{Chiba University, Chiba}
\affiliation{Chonnam National University, Kwangju}
\affiliation{University of Cincinnati, Cincinnati, Ohio 45221}
\affiliation{University of Hawaii, Honolulu, Hawaii 96822}
\affiliation{High Energy Accelerator Research Organization (KEK), Tsukuba}
\affiliation{Hiroshima Institute of Technology, Hiroshima}
\affiliation{Institute of High Energy Physics, Chinese Academy of Sciences, Beijing}
\affiliation{Institute of High Energy Physics, Vienna}
\affiliation{Institute for Theoretical and Experimental Physics, Moscow}
\affiliation{J. Stefan Institute, Ljubljana}
\affiliation{Kanagawa University, Yokohama}
\affiliation{Korea University, Seoul}
\affiliation{Swiss Federal Institute of Technology of Lausanne, EPFL, Lausanne}
\affiliation{University of Ljubljana, Ljubljana}
\affiliation{University of Maribor, Maribor}
\affiliation{University of Melbourne, Victoria}
\affiliation{Nagoya University, Nagoya}
\affiliation{Nara Women's University, Nara}
\affiliation{National Central University, Chung-li}
\affiliation{National United University, Miao Li}
\affiliation{Department of Physics, National Taiwan University, Taipei}
\affiliation{H. Niewodniczanski Institute of Nuclear Physics, Krakow}
\affiliation{Nippon Dental University, Niigata}
\affiliation{Niigata University, Niigata}
\affiliation{Nova Gorica Polytechnic, Nova Gorica}
\affiliation{Osaka City University, Osaka}
\affiliation{Osaka University, Osaka}
\affiliation{Panjab University, Chandigarh}
\affiliation{Peking University, Beijing}
\affiliation{Princeton University, Princeton, New Jersey 08544}
\affiliation{University of Science and Technology of China, Hefei}
\affiliation{Seoul National University, Seoul}
\affiliation{Shinshu University, Nagano}
\affiliation{Sungkyunkwan University, Suwon}
\affiliation{University of Sydney, Sydney NSW}
\affiliation{Tata Institute of Fundamental Research, Bombay}
\affiliation{Toho University, Funabashi}
\affiliation{Tohoku Gakuin University, Tagajo}
\affiliation{Tohoku University, Sendai}
\affiliation{Department of Physics, University of Tokyo, Tokyo}
\affiliation{Tokyo Institute of Technology, Tokyo}
\affiliation{Tokyo Metropolitan University, Tokyo}
\affiliation{Tokyo University of Agriculture and Technology, Tokyo}
\affiliation{University of Tsukuba, Tsukuba}
\affiliation{Virginia Polytechnic Institute and State University, Blacksburg, Virginia 24061}
\affiliation{Yonsei University, Seoul}
   \author{Q.~L.~Xie}\altaffiliation{Present adderess: Physics Department of Sichuan University, Sichuan Province, China} \affiliation{Institute of High Energy Physics, Chinese Academy of Sciences, Beijing}
   \author{K.~Abe}\affiliation{High Energy Accelerator Research Organization (KEK), Tsukuba} 
   \author{K.~Abe}\affiliation{Tohoku Gakuin University, Tagajo} 
   \author{I.~Adachi}\affiliation{High Energy Accelerator Research Organization (KEK), Tsukuba} 
   \author{H.~Aihara}\affiliation{Department of Physics, University of Tokyo, Tokyo} 
   \author{Y.~Asano}\affiliation{University of Tsukuba, Tsukuba} 
   \author{T.~Aushev}\affiliation{Institute for Theoretical and Experimental Physics, Moscow} 
   \author{S.~Bahinipati}\affiliation{University of Cincinnati, Cincinnati, Ohio 45221} 
   \author{A.~M.~Bakich}\affiliation{University of Sydney, Sydney NSW} 
   \author{E.~Barberio}\affiliation{University of Melbourne, Victoria} 
% ...
\collaboration{The Belle Collaboration}

\date{\textbf{\today}}

\begin{abstract}
 We report the observation of $B^{-} \to J/\psi \Lambda \bar{p}$ and 
searches for $B^{-} \to J/\psi \Sigma^{0} \bar{p}$ and  
$B^{0} \to J/\psi p \bar{p}$ decays,
using a sample of 275 million $B\bar{B}$ pairs collected 
with the Belle detector at the $\Upsilon(4S)$ resonance.
We observe a signal of $17.2 \pm 4.1$ events 
with a significance of 11.1$\sigma$
and obtain a branching fraction of
 $\mathcal{B}(B^{-}\to J/\psi \Lambda \bar{p})$ =
 $11.6\pm2.8({\rm stat.})_{-2.3}^{+1.8}({\rm sys.}) \times 10^{-6}$. 
No signal is found for either of the two decay modes, 
$B^{-}\to J/\psi \Sigma^{0} \bar{p}$ and  $B^{0}\to J/\psi p \bar{p}$, 
and upper limits for the branching fractions are determined to be 
$\mathcal{B}(B^{-}\to J/\psi \Sigma^{0} \bar{p}) < 1.1 \times 10^{-5}$ and 
$\mathcal{B}(B^{0}\to J/\psi p \bar{p}) < 8.3 \times 10^{-7}$ 
at 90\% confidence level.
\end{abstract}

 \maketitle

We report the observation of the decay mode $B^-\to J/\psi \Lambda \bar{p}$,
which is a new type of baryonic $B$ decay, $B \to$ charmonium + baryons. 
Evidence for this mode was reported 
in previous studies by  
BaBar \cite{4} and Belle \cite{zsl}, with 89 million and 85 million 
$B\bar{B}$ pairs, respectively.

Originally, modes of this type were proposed as a potential explanation  
\cite{BN} for the excess in the low momentum region
of the inclusive $J/\psi$ momentum spectrum for $B\to J/\psi + X$ \cite{1,2}, 
which is not consistent with the predictions from nonrelativistic QCD 
calculations \cite{NRQCD}. 
Although the measured branching fraction of 
$B^-\to J/\psi \Lambda \bar{p}$, ${\cal O}(10^{-5})$, 
is not large enough to explain the observed excess, 
this result stimulated experiments to explore 
new baryonic $B$ decay modes in addition to 
those that had already been observed, namely: 
$B \to$ charmed baryon (+ light anti-baryon and mesons) \cite{CBNM}, 
$B \to D$ + baryons \cite{DNN},
$B \to$ charmless baryons + light mesons \cite{NNM}, 
and $B \to $ baryons + $\gamma$ \cite{NNG}. 
One of the interesting features of many three-body baryonic 
$B$ decays is the presence of an enhancement 
in the threshold region of the baryon anti-baryon 
invariant mass spectrum. 
Further studies of $B \to J/\psi \Lambda \bar p$ and related baryonic 
decay modes are expected to provide additional 
information on the baryon formation mechanism in $B$ decays.

In this paper, we report the observation of $B^{-} \to J/\psi \Lambda\bar{p}$
and searches for $B^{-} \to J/\psi \Sigma^{0} \bar{p}$ and  
$B^{0} \to J/\psi p \bar{p}$ \cite{CC} in a 
sample of 253 $\rm fb^{-1}$ containing 275 million $B\bar{B}$ pairs
accumulated at the $\Upsilon$(4$S$) resonance with the Belle 
detector \cite{belle} at the KEKB energy 
asymmetric $e^+e^-$ collider \cite{kekb}.

The Belle detector is a large-solid-angle magnetic spectrometer that consists
of a silicon vertex detector (SVD), a 50-layer central drift
chamber (CDC), an array of aerogel threshold \v{C}erenkov counters (ACC), a
barrel-like arrangement of time-of-flight scintillation counters (TOF), and an
electromagnetic calorimeter comprised of CsI(Tl) crystals (ECL). These
detectors are located inside a superconducting solenoid coil that provides a
1.5 T magnetic field. An iron flux-return located outside of the coil is
instrumented to detect $K_L$ mesons and to identify muons (KLM). 

Candidates for $B^{-} \to J/\psi \Lambda \bar{p}$ and 
$B^{0} \to J/\psi p \bar{p}$
are fully reconstructed 
with all daughters in the final states, 
where we use the decay chains $J/\psi \to l^+ l^-$ ($l = e, \mu$) 
and $\Lambda \to p \pi^-$.
Candidates for $B^{-} \to J/\psi \Sigma^0 \bar{p}$ are
reconstructed partially, where only the daughters from the 
$J/\psi$ and $\Lambda$ decays plus the 
anti-proton are reconstructed in the final state. 
In this case, the $\Lambda$ originates from 
the $\Sigma^0 \to \Lambda \gamma$ decay. 
Partial reconstruction of $B^{-} \to J/\psi \Sigma^0 \bar{p}$ 
without the low momentum $\gamma$ from the 
$\Sigma^0$ decay gives better sensitivity
than full reconstruction 
because of the low detection efficiency for
soft photons. As a result, the same selection criteria are
applied for fully reconstructed $B^{-} \to J/\psi \Lambda \bar{p}$ and
partially reconstructed $B^{-} \to J/\psi \Sigma^0 \bar{p}$.
  
Events are required to pass a basic hadronic event selection \cite{PRD03}.
To suppress continuum background 
($e^+e^- \to q \bar q$, where $q = u, d, s, c$),  
we require the ratio of the second to zeroth Fox-Wolfram 
moments \cite{FW} to be less than 0.5.

The selection criteria for the $J/\psi$ 
decaying to $l^+l^-$ 
are identical to those used in
our previous papers \cite{zsl,PRD03}. 
 $J/\psi$ candidates are pairs of oppositely 
charged tracks that originate from within 5~cm 
of the nominal interaction point (IP) along 
the beam direction and are positively identified as leptons.
In order to reduce the effect of
bremsstrahlung or final state radiation, photons detected in the ECL within
$0.05$ radians of the original $e^-$ or $e^+$ direction are included in the
calculation of the $e^+e^-(\gamma)$ invariant mass. 
Because of the radiative low-mass tail,
the $J/\psi$ candidates are required 
to be within an asymmetric invariant mass window: $-150(-60)$ $\rm{MeV}$/$c^2$
$<M_{e^+e^-(\gamma)}(M_{\mu^+\mu^-})-m_{J/\psi}
<$+36(+36) $\rm{MeV}$/$c^2$, where $m_{J/\psi}$ is the nominal 
$J/\psi$ mass \cite{pdg}.
In order to improve the momentum resolution of the $J/\psi$ signal, 
a vertex and mass constrained fit to the reconstructed $J/\psi$ candidates
is then performed and a loose cut on the vertex fit quality is applied.

In order to identify hadrons, a likelihood $L_i$ 
for each  hadron type $i$ ($i = \pi, K$ and $p$) is formed 
using information from the ACC, TOF and $dE/dx$. 
Charged tracks that were previously identified as electrons or muons 
are rejected in the hadron identification procedure. 
The proton and pion from the $\Lambda$ decay are selected with 
the requirements of $L_{p}/(L_{p}+L_{K})> 0.6$ 
and $L_{\pi}/(L_{\pi}+L_{K})> 0.6$, 
which have efficiencies of 94.8\% and 99.2\%, respectively.

The primary protons from the $B^-$ ($B^{0}$) decay are selected
with requirements of 
$L_{p}/(L_{p}+L_{K})> 0.6~(0.7) $ and 
$L_{p}/(L_{p}+L_{\pi})> 0.9~(0.9)$. 
For the primary proton, we also apply 
requirements on $dr$ and $dz$, the impact parameters 
perpendicular to and along the beam direction with 
respect to the IP, respectively. 
The efficiency of $dr$ and $dz$ requirements is 97.8\% (91.2\%) 
for $B^-$ ($B^0$) signal modes. 

For $\Lambda$ candidates, 
we impose momentum-dependent requirements on the $dr$ 
values for both $\Lambda$ daughter tracks,
the distance between the daughter tracks along the beam axis at the
$\Lambda$ vertex, the difference of azimuthal angles between the $\Lambda$ 
momentum and the direction of the $\Lambda$ vertex from the IP,
and the flight length of the $\Lambda$, 
which retain 81.11\% of $\Lambda$ candidates. 
The invariant mass of the $\Lambda$ candidate is required 
to be within 5~$\rm{MeV}/c^2$ ($3\sigma$) of the $\Lambda$ mass. 
We apply vertex and mass constrained fits for the $\Lambda$
candidates in order to improve the momentum resolution.

$B$ mesons are reconstructed by combining a $J/\psi$, a $\Lambda(p)$ 
and a $\bar{p}$ candidate.
To reduce combinatorial background, we impose a requirement on the
quality ($\chi^2$) of the vertex fit for the leptons from $J/\psi$
and primary proton(s) that retains 95.8\% (95.1\%) of 
$B^-$ ($B^0$) signal.

These criteria maximize 
$N_{\rm sig}/\sqrt{N_{\rm sig}+N_{\rm bkg}}$, 
where $N_{\rm sig}$ is the number of expected signal events from signal 
Monte Carlo (MC) samples with assumed branching fractions of 
$1.0 \times 10^{-5}$ for $B^{-} \to J/\psi \Lambda \bar{p}$ 
and $1.0 \times10^{-6}$ for $B^{0} \to J/\psi p \bar{p}$, 
and $N_{\rm bkg}$ is the number of expected background 
events estimated from the sideband data.

We identify $B$ candidates using two kinematic 
variables calculated in the center-of-mass system: the beam-energy 
constrained mass ($\Mbc \equiv \sqrt{E_{\rm beam}^{2}-P_{B}^{2}}$) and 
the mass difference ($\Dmb \equiv M_{B}- m_{B}$) \cite{dmb}, 
where $E_{\rm{beam}}$ is the beam energy, $P_{B}$ and $M_{B}$ are the 
reconstructed momentum and mass of the $B$ candidate, 
and $m_{B}$ is the nominal $B$ mass. 
$B^-$ ($B^0$) candidates within $|\Dmb|<0.20$ $\rm{GeV}/c^{2}$ 
($-0.32$ $\rm{GeV}/c^{2}$ $<\Dmb<0.20$ $\rm{GeV}/c^{2}$) and 
$\Mbc >5.20$ $\rm{GeV}/c^{2}$ 
are selected for the final analysis. The signal regions
are defined to be $5.27$ GeV/$c^2$ $< \Mbc < 5.29$ GeV/$c^2$ and 
$|\Dmb| <$ 0.03 ${\rm GeV}/c^2$ for $B^{-} \to J/\psi \Lambda \bar{p}$, 
$5.265$ ${\rm GeV}/c^2$ $< \Mbc <$ $5.29$ ${\rm GeV}/c^2$ and 
$-0.104$ ${\rm GeV}/c^2$ $< \Dmb < -0.044$ ${\rm GeV}/c^2$ for partially 
reconstructed $B^{-} \to J/\psi \Sigma^0 \bar{p}$, 
and $5.27$ ${\rm GeV}/c^2$ $< \Mbc <$ $5.29$ ${\rm GeV}/c^2$ and 
$|\Dmb| < 0.02$ ${\rm GeV}$/$c^2$ for $B^{0} \to J/\psi p \bar{p}$, 
which corresponds to three standard deviations based on the MC simulation. 
The candidates outside of the signal region are used to study background.

After all selection requirements, about 11\% of the 
$B^{-} \to J/\psi \Lambda \bar{p}$ candidates have 
more than one entry per event; this occurs for 20\% of the 
partially reconstructed $B^{-} \to J/\psi \Sigma^0 \bar{p}$ 
and 3.8\% of the $B^{0} \to J/\psi p \bar{p}$ candidates. 
For these events, the $B$ candidate with the best vertex quality is selected. 
If an event satisfies both $B^{-} \to J/\psi \Lambda \bar{p}$ and 
$B^{0} \to J/\psi p \bar{p}$ selection criteria, we take it only as 
a $B^{-} \to J/\psi \Lambda\bar{p}$ candidate.

The signal yields are extracted by maximizing the two dimensional (2D)
extended likelihood function,
\[
{\cal L} = \frac{e^{-\sum
\limits_{k}N_{k}}}{N!}\prod^N_{i=1}\left[\sum_{k}N_{
k}\times
P_k(M_{{\rm bc}}^{i},\Delta M_B^{i})\right],
\] 
where $N$ is the total number of candidate events,
$i$ is the identifier of the $i$-th event,
$N_{k}$ and $P_{k}$ are the
yield and  probability density function (PDF) of the component $k$,
which corresponds to the signal and background. 
The signal yields for $B^{-} \to J/\psi \Lambda \bar{p}$ and partially 
reconstructed $B^{-} \to J/\psi \Sigma^0 \bar{p}$ are simultaneously 
extracted from the $B^{-} \to J/\psi \Lambda \bar{p}$ candidate sample.

The signal PDFs for all three decay modes are modeled 
using a sum of two Gaussians
for $\Mbc$ and a sum of two Gaussians plus a bifurcated Gaussian 
that describes the tail of the distribution 
for $\Dmb$. 
The PDF parameters are initially determined using signal MC; 
the primary Gaussian parameters are then corrected using the 
$B^+ \to J/\psi K^{*+} (K^{*+} \to K_s \pi^+)$ control sample. 
The parameters are kept fixed in the fits to the data.

The dominant background comes from random combinations of $J/\psi$, 
$\Lambda$ ($p$), and $\bar p$ candidates.
A threshold function \cite{ARGUS} is used for the 
$\Mbc$ PDF. For $\Dmb$, we take the effect of the 
kinematic boundary into account by using the function
\[
     P_{\rm thr}(x; x_c,p,c) = \left\{
   \begin{array}{ll}
     (x - x_c)^p e^{-c(x-x_c)} & (x \ge x_c) \\
     0                         & (x < x_c) \\
   \end{array}\right.
\]
with $x = \Dmb$ and 
$ x_c= - (m_{B}-m_{J/\psi}-m_{\Lambda}(m_{p})-m_{\bar{p}})$, 
where $m_{B}$, $m_{J/\psi}$,  $m_{\Lambda}$ ($m_{p}$) and $m_{\bar{p}}$ 
are the nominal masses of the $B$, $J/\psi$, $\Lambda$ ($p$) 
and ${\bar{p}}$, respectively.

\begin{table*}[!hbtp]
\begin{center}
\caption{\label{result} Summary of the results. $Y$ and $b$ are the fitted 
signal and total background yields in the signal region, 
$n_{0}$ is the observed number of candidate events in the signal region, $\epsilon$ (error includes systematic error) is the detection efficiency, $Y_{90}$ is the upper limit on the signal yield at 90\% confidence level and ${\cal B}$ is the branching fraction. 
}
\begin{tabular}{ccccccc}\hline \hline
mode & $Y$ &$b$& $n_0$& $\epsilon(\%)$ &$Y_{90}$& ${\cal B}$\\
\hline
$B^{-} \to J/\psi \Lambda \bar{p}$ & $17.2\pm 4.1$  
 & $0.41\pm 0.09({\rm stat.})$&16&$7.2_{-1.4}^{+1.1}$
 & $-$&$11.6\pm2.8({\rm stat.})_{-2.3}^{+1.8}({\rm sys.}) \times 10^{-6}$ \\
$B^{-} \to J/\psi \Sigma^0 \bar{p}$  &$-1.1\pm 1.7$
 & $0.31\pm 0.04({\rm stat.})\pm 0.03({\rm sys.})$& 1&$2.3^{+0.9}_{-0.8}$ 
 & $<5.3$ &$<1.1 \times 10^{-5}$\\
$B^{0} \to J/\psi p \bar{p}$  &$-6.1\pm 2.2$
 & $0.94\pm 0.10({\rm stat.})_{-0.16}^{+0.04}({\rm sys.})$& 3
 & $26.4^{+6.8}_{-5.4}$ & $<7.1$ &$<8.3 \times 10^{-7}$\\
\hline\hline
\end{tabular}
\end{center}
\end{table*}

\begin{figure*}[hbtp]
\includegraphics[width=0.3\textwidth]{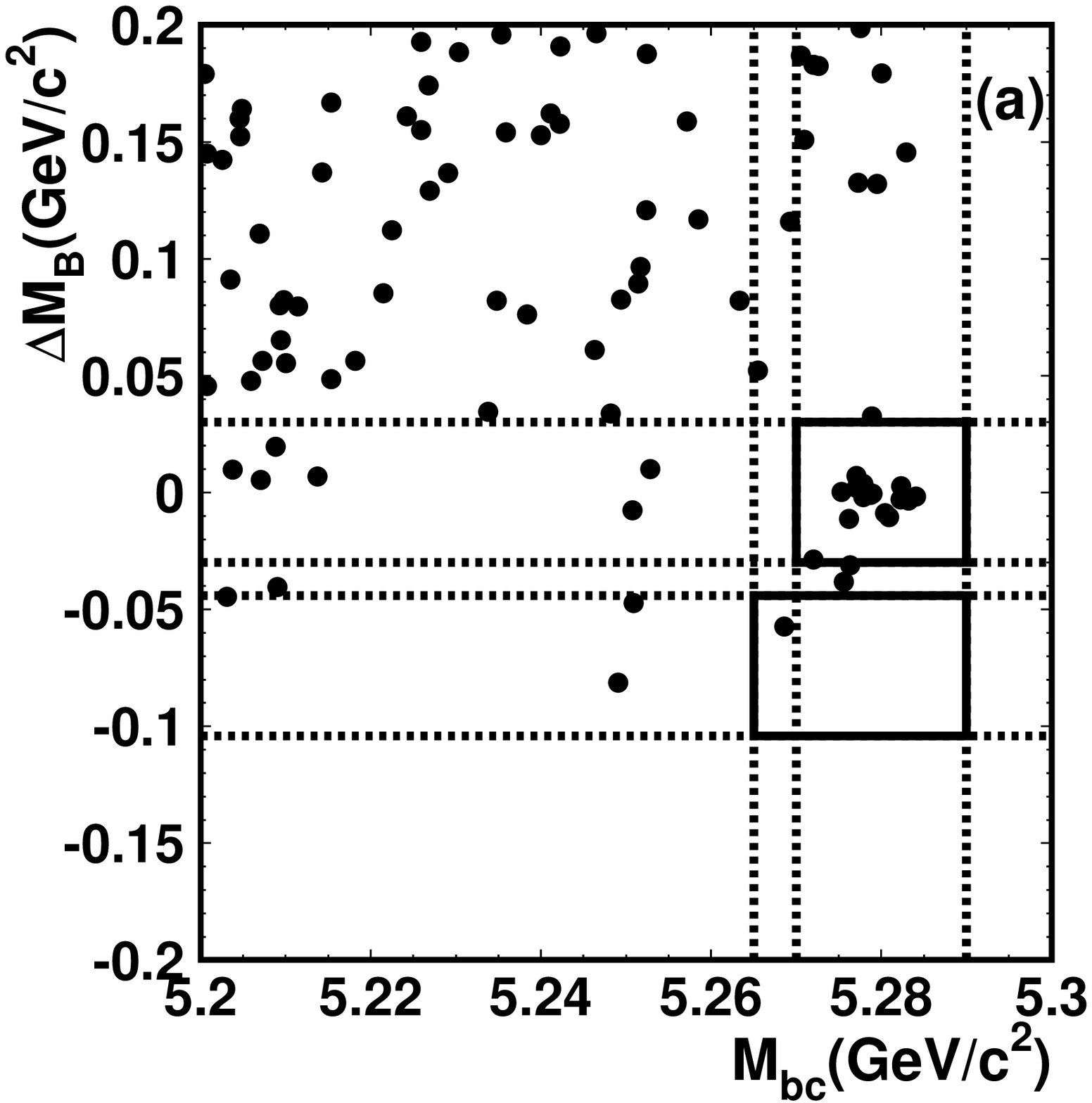}
\includegraphics[width=0.3\textwidth]{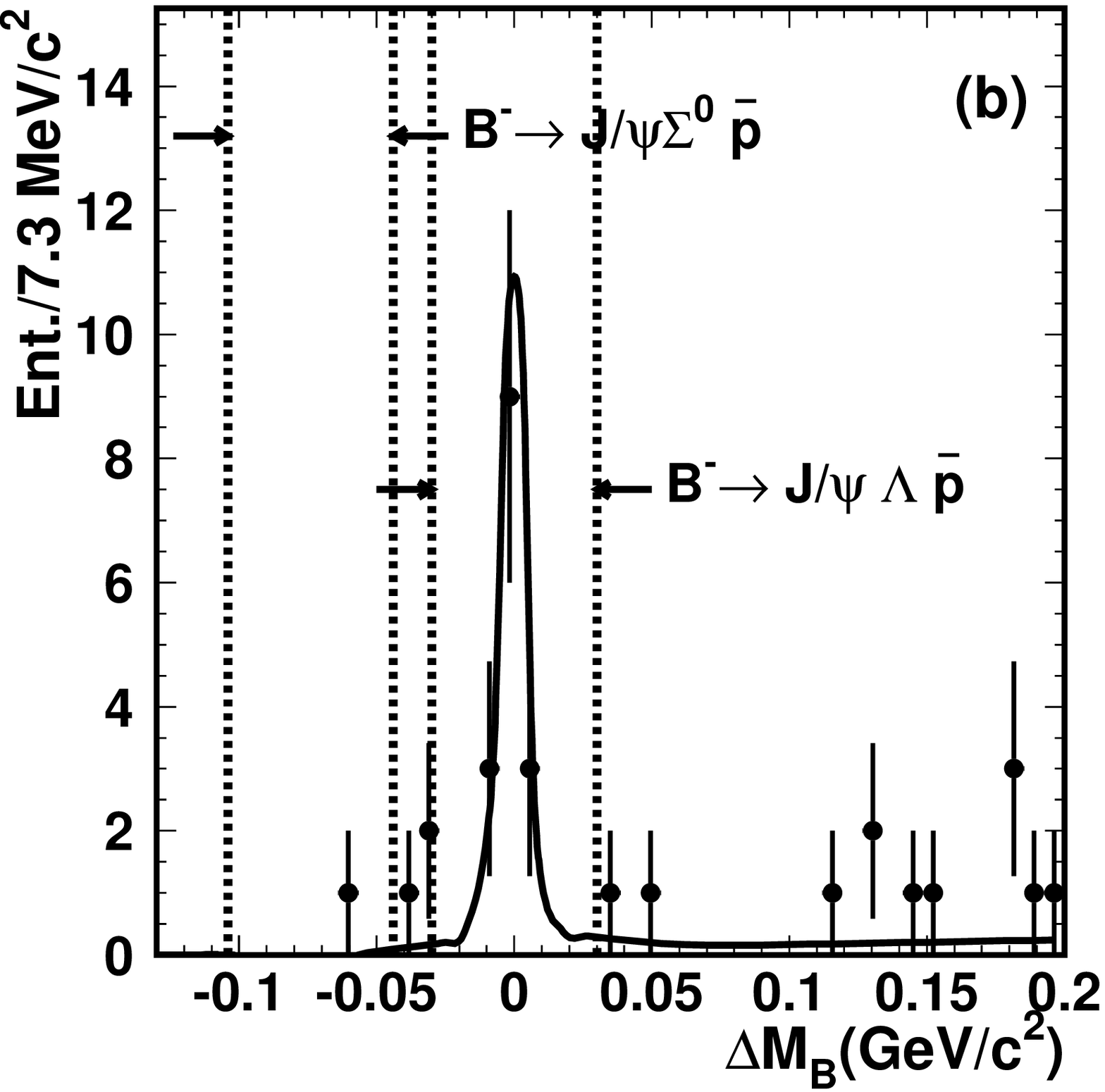}
\includegraphics[width=0.3\textwidth]{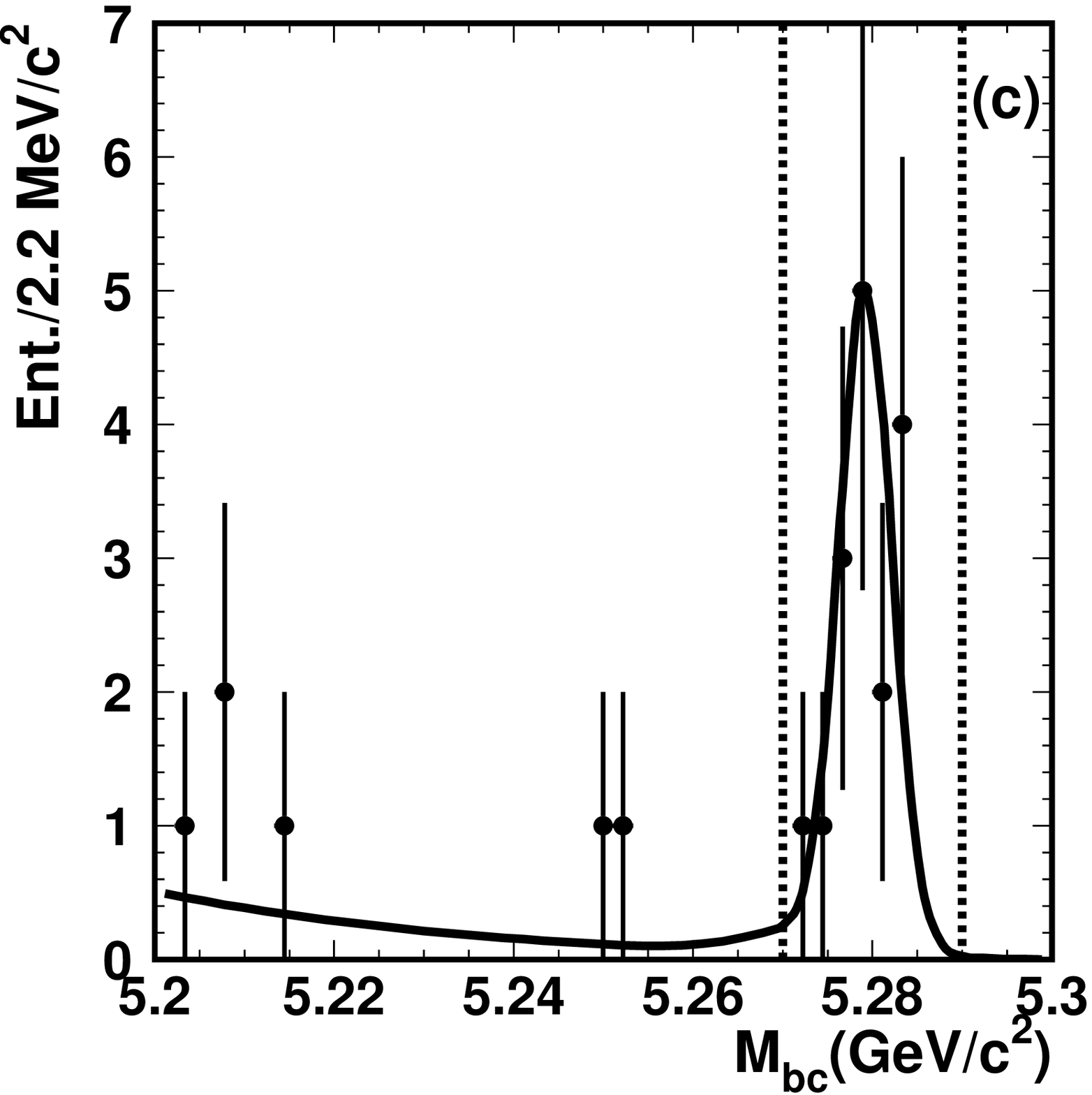}
\caption{
(a) The ($\Mbc$, $\Dmb$) scatterplot of $B^{-} \to J/\psi \Lambda \bar{p}$ 
candidates and its projections onto 
(b) $\Dmb$ with 5.265 GeV/$c^2$ $< \Mbc <$ 5.29 GeV/$c^2$ and 
(c) $\Mbc$ with $ |\Dmb| < 0.03$ GeV/$c^2$.
The dashed lines and solid boxes indicate the signal regions. The curves are
the result of the fit as described in the text. }
\label{sbox}
\end{figure*}

\begin{figure*}[htp]
\includegraphics[width=0.3\textwidth]{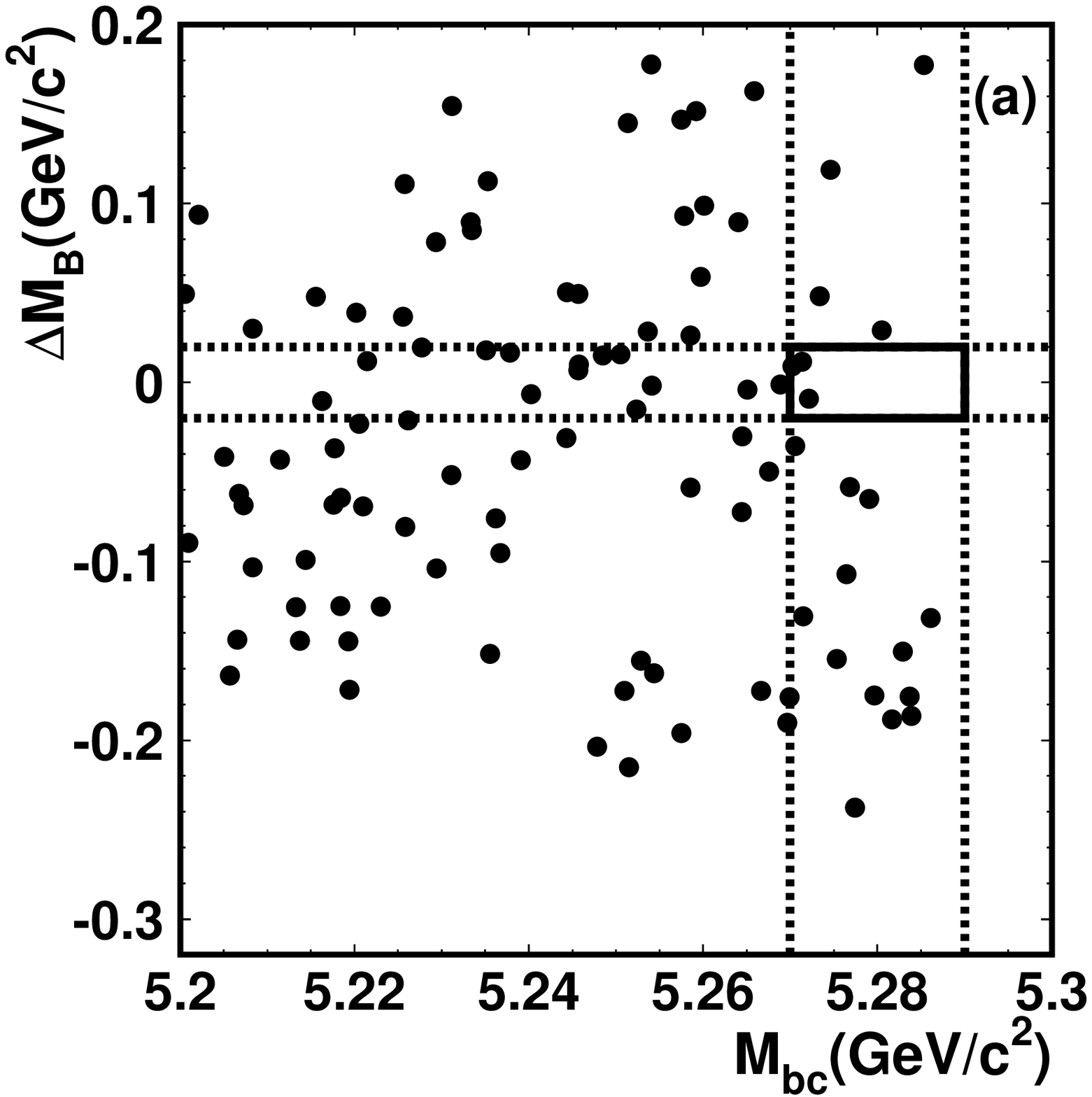}
\includegraphics[width=0.3\textwidth]{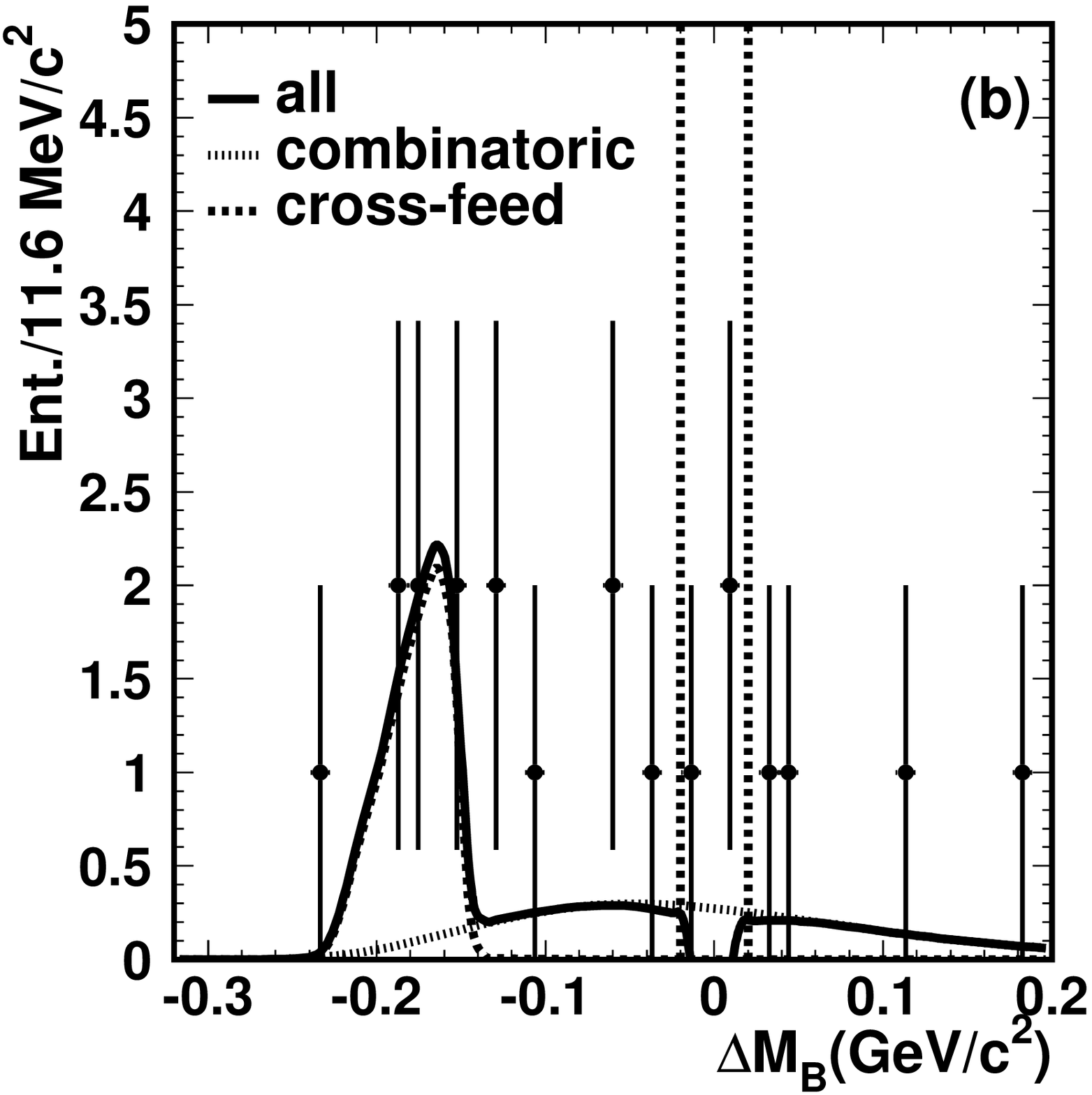}
\includegraphics[width=0.3\textwidth]{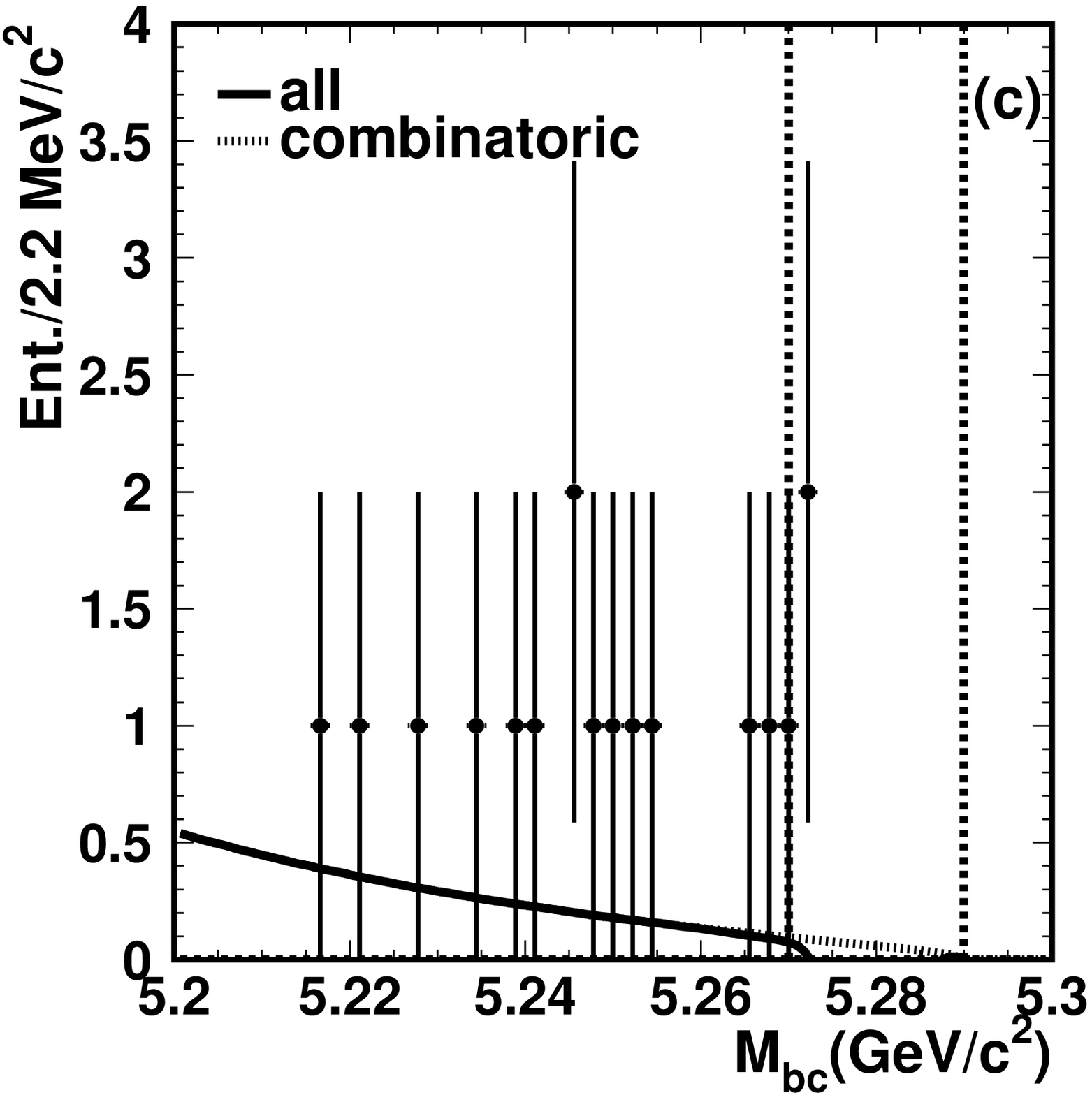}
\caption{
(a) ($\Mbc$, $\Dmb$) scatterplot of $B^{0} \to J/\psi p \bar{p}$ 
candidates and its projections on to 
(b) $\Dmb$ with 5.27 GeV/$c^2$ $< Mbc <$ 5.29 GeV/$c^2$ and
(c) $\Mbc$ with $ |\Dmb| < 0.02$ GeV/$c^2$.
The dashed lines and the solid box indicate the signal regions. The curves are
the result of the fit. }
\label{sbox2}
\end{figure*}

For $B^{0} \to J/\psi p \bar{p}$,
a cross-feed from $B^{-} \to J/\psi \Lambda \bar{p}$,
where the $\pi$ from the $\Lambda$ is undetected, forms 
a peak in the negative $\Dmb$ sideband region.
The PDF for this cross-feed is modeled by a smoothed
histogram from the $B^{-} \to J/\psi \Lambda \bar{p}$ MC sample.

In the fit, the value of $N_k$ and the parameters for combinatoric
background are allowed to float.
Figures \ref{sbox} and \ref{sbox2} show the ($\Mbc$, $\Dmb$) scatterplots and
their projections for candidates after all selections are applied. 
The fit results are superimposed on the projections.
There are sixteen candidate events in the signal region for
$B^{-} \to J/\psi \Lambda \bar{p}$, one for 
$B^{-} \to J/\psi \Sigma^0 \bar{p}$ and three for $B^{0} \to J/\psi p \bar{p}$.

Table~\ref{result} summarizes the maximum-likelihood 
fit results for the signal ($Y$) and signal-region background ($b$) 
yields and their statistical errors.
For the $B^{-}\to J/\psi \Lambda \bar{p}$ decay the fit gives $17.2 \pm 4.1$ 
signal events with a statistical significance of $11.1 \sigma$. 
The statistical significance is defined as 
$\sqrt{-2\ln({\cal L}_{0}/{\cal L}_{\rm max})}$, 
where ${\cal L}_{\rm max}$ and ${\cal L}_{0}$ denote the maximum likelihood with the
fitted signal yield and with the yield fixed at zero, respectively.
No significant signal is found for the $B^{-}\to J/\psi
\Sigma^{0} \bar{p}$ and  $B^{0}\to J/\psi p \bar{p}$ decay modes,
while the number of cross-feed events in the $\Dmb$ sideband of 
$B^{0}\to J/\psi p \bar{p}$ ($9.0 \pm 3.2$) is consistent with the
expectation from the observed 
$B^{-}\to J/\psi \Lambda \bar{p}$ signal yield ($7.0 \pm 1.7$).
For these two modes, 
we obtain upper limits on the yield at 90\% 
confidence level ($Y_{90}$) using the
Feldman-Cousins method \cite{FC}, which takes into account the systematic 
errors due to the uncertainties in the signal detection 
efficiency and the background yield \cite{Conrad}.

 The branching fraction is determined with  
 $N_{S}/[\epsilon \times N_{B\bar{B}} \times {\cal B}(J/\psi \to l^{+}l^{-})
  \times {\cal B}(\Lambda \to p \pi^{-})]$ 
for $B^{-}\to J/\psi \Lambda \bar{p}$ and the partially reconstructed 
$B^{-}\to J/\psi \Sigma^{0} \bar{p}$. 
For $B^{0}\to J/\psi p \bar{p}$ we use 
$N_{S}/[\epsilon \times N_{B\bar{B}}\times {\cal B}(J/\psi \to l^{+}l^{-})]$.
Here $N_{S}$ is the signal yield, $N_{B\bar{B}}$ is the number of 
$B\bar{B}$ pairs. 
We use the world averages \cite{pdg} for the branching fractions of 
${\cal B}(J/\psi \to l^{+}l^{-})$ and  ${\cal B}(\Lambda \to p \pi^{-})$.
The efficiencies ($\epsilon$) are
determined from the signal MC sample with the same selection as 
used in the data. 
A three-body phase space model is employed for all three decay modes. 
The fractions of neutral and charged $B$ mesons produced in $\Upsilon(4S)$ 
decays are assumed to be equal.

\begin{figure*}[htp]
    \includegraphics[width=.8\textwidth]{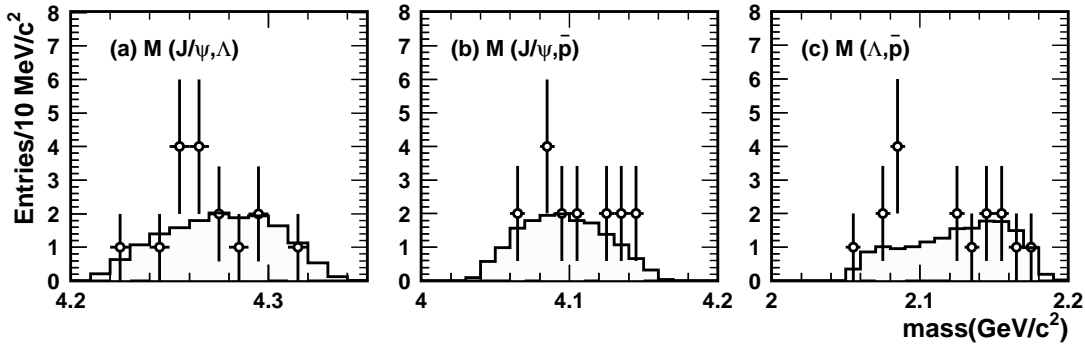}
   \caption{ 
    Invariant mass distributions for (a) $M(J/\psi \Lambda)$, 
    (b) $M(J/\psi \bar p)$ and (c) $M(\Lambda \bar p)$ 
    for the sixteen signal-region $B^- \to J/\psi \Lambda \bar{p}$ candidates. 
    The histograms are phase space distributions from the signal MC sample
     normalized to sixteen events.}
     \label{mom}
  \end{figure*}

Figure~\ref{mom} demonstrates the consistency between 
the observed invariant mass distributions $M(J/\psi \Lambda)$, 
$M(J/\psi \bar p)$, and $M(\Lambda \bar p)$ of the sixteen 
$B^- \to J/\psi \Lambda \bar{p}$ candidates and the phase 
space distributions obtained from signal MC.

The sources and sizes of systematic uncertainties are summarized in
Table~\ref{sys-err}. 
The systematic uncertainty on the yield is examined  %YS determined 
by varying each fixed shape parameter by $\pm1\sigma$. 
A possible bias in the fitting is studied using MC samples; 
no significant bias is found. 
The systematic uncertainty assigned to the yield is 2.5\%.

The dominant sources of systematic errors on the efficiency are 
uncertainties in the three-body decay model, $\Lambda$ reconstruction, 
tracking efficiency and particle identification. 
To estimate the error due to uncertainty in decay modeling
for $B^- \to J/\psi \Lambda \bar p$, we subdivide phase space
into a few bins and recompute the branching fraction using
MC-determined bin-by-bin efficiencies. The maximum difference between
the nominal value of the branching fraction and the values
obtained with different choices of bins is assigned as a
systematic error.
For $B^- \to J/\psi \Sigma^0 \bar{p}$ ($B^0 \to J/\psi p \bar{p}$), 
the distribution in phase space is unknown and we conservatively 
assign the maximum variation of efficiency among the slices of 
M($J/\psi$,$\Lambda$), M($J/\psi$,$\bar{p}$) and M($\Lambda$,$\bar{p}$) 
[M($J/\psi$,$p$), M($J/\psi$,$\bar{p}$) and M($p$,$\bar{p}$)] 
as a systematic uncertainty. 
The $\Lambda$ reconstruction error is determined by 
comparing the transverse proper 
flight distance distributions for data and MC simulation. 
The discrepancies weighted by distributions of the $\Lambda$ momentum and 
the decay length are assigned as systematic errors. 
The uncertainties in the tracking efficiency 
are estimated by linearly adding the momentum-dependent 
single track systematic errors. 
We assign uncertainties of 3\% per proton, 2\% per pion, 
and 2\% per lepton for the particle and lepton identification.

The systematic errors for the background yield are evaluated 
by varying each of the PDF parameters by its  
statistical error from the fit and by fixing the signal 
yield at zero in the case of $B^{0}\to J/\psi p \bar{p}$ 
to accommodate the possibility that an overestimate 
of the background might be causing 
the large negative signal yield ($-6.1 \pm 2.2$).

\begin {table*}[htp]
\begin {center}
\caption {Summary of systematic uncertainties (\%) in yield and detection efficiency.}
\begin {tabular}{cccc}
\hline
\hline
Source&$B^{-}\to J/\psi \Lambda \bar{p}$ &\ \  $B^{-}\to J/\psi \Sigma^0 \bar{p}$ &\ \  $B^{0}\to J/\psi p \bar{p}$\\
\hline
Uncertainty in yield & $\pm 2.5$ & - & -  \\
Tracking Error & $\pm 8.6$ &$\pm 9.8$ & $\pm 4.8$  \\
PID(proton and pion) & $\pm 8.0$ & $\pm 8.0$ & $\pm 6.0$ \\
Lepton ID & $\pm 4.0$ & $\pm 4.0$ & $\pm 4.0$ \\
$\Lambda$ Reconstruction & $\pm 7.9$ & $\pm 9.8$ & -\\
$\Lambda$ BR & $\pm 0.8$ & $\pm 0.8$ & - \\
MC Statistics & $\pm 0.2$ & $\pm 0.2 $ & $\pm 0.5$ \\
3-body decay model & $+4.7$/$-12.6$ & $+38.0$/$-29.1$ & $+24.1$/$-18.8$   \\
\hline
Total & $+15.7$/$-19.6$ & $+41.4$/$-33.4$ & $+25.6$/$-20.7$ \\
\hline
\hline
\end {tabular}
\label{sys-err}
\end {center}
\end {table*}

In summary, 
we have observed a $17.2 \pm 4.1$ event signal  
for $B^{-}\to J/\psi\Lambda \bar{p}$, 
with a statistical significance of $11.1 \sigma$.
The measured branching fraction is  
$\mathcal{B}$($B^- \to J/\psi \Lambda \bar{p}$) 
= $11.6\pm2.8({\rm stat.})_{-2.3}^{+1.8}({\rm sys.}) \times 10^{-6}$. 
This establishes a new type of baryonic $B$ decay, $B \to$ charmonium +
baryon anti-baryon. The $\Lambda \bar{p}$ distribution for this decay is
consistent with a phase space, in contrast to 
$B \to \Lambda \bar{p} \pi$ \cite{NNM} and
other baryonic B decays.
No significant signals are found for the 
$B^{-}\to J/\psi \Sigma^{0} \bar{p}$ and  
$B^{0}\to J/\psi p \bar{p}$ decay modes. 
We obtain upper limits on the branching fractions of 
$\mathcal{B}(B^{-}\to J/\psi \Sigma^{0} \bar{p}) < 1.1 \times 10^{-5}$ 
and $\mathcal{B}(B^{0}\to J/\psi p \bar{p}) < 8.3 \times 10^{-7}$ 
at 90\% confidence level. 

\begin{acknowledgments}
We thank the KEKB group for the excellent operation of the
accelerator, the KEK cryogenics group for the efficient
operation of the solenoid, and the KEK computer group and
the NII for valuable computing and Super-SINET network
support.  We acknowledge support from MEXT and JSPS (Japan);
ARC and DEST (Australia); NSFC (contract No.~10175071,
China); DST (India); the BK21 program of MOEHRD, and the
CHEP SRC and BR (grant No. R01-2005-000-10089-0) programs of
KOSEF (Korea); KBN (contract No.~2P03B 01324, Poland); MIST
(Russia); MHEST (Slovenia);  SNSF (Switzerland); NSC and MOE
(Taiwan); and DOE (USA).
\end{acknowledgments}


\begin{thebibliography}{999}
\bibitem{4} Babar Collaboration, B. Aubert \emph{et al.},  Phys. Rev. Lett.
  \textbf{90}, 231801 (2003).
\bibitem{zsl} Belle Collaboration, S.L.~Zang \emph{et al.}, Phys. Rev. D
  \textbf{69}, 017101 (2004).
\bibitem{BN} S.J.~Brodsky and F.S.~Navarra, Phys. Lett. B
   \textbf{411}, 152 (1997).
\bibitem{1} CLEO Collaboration, R. Balest \emph{et al.}, Phys. Rev. D{\bf 52},
  2661(1995);
  S.~Anderson \emph{et al.}, Phys. Rev. Lett.  \textbf{89}, 282001 (2003).
\bibitem{2} Babar Collaboration, B. Aubert \emph{et al.}, Phys. Rev. D
  \textbf{67}, 032002 (2003).
\bibitem{NRQCD} M.~Beneke, G.A.~Schuler, and S.~Wolf, Phys. Rev. D{\bf 62},
  034004 (2003).
\bibitem{CBNM} CLEO Collaboration, X.~Fu \emph{et al.}, 
  Phys. Rev. Lett. \textbf{79}, 3125 (1997);
  Belle Collaboration, N.~Gabyshev \emph{et al.}, 
   Phys. Rev. D  \textbf{66}, 091102(R) (2002);
  Belle Collaboration, N.~Gabyshev \emph{et al.}, 
   Phys. Rev. Lett.  \textbf{90}, 121802 (2003);
\bibitem{DNN} CLEO Collaboration, S.~Anderson \emph{et al.}, 
  Phys. Rev. Lett. \textbf{86}, 2732 (2001);
  Belle Collaboration, K.~Abe \emph{et al.}, Phys. Rev. Lett.
  \textbf{89}, 151802 (2002).
\bibitem{NNM} Belle Collaboration, K.~Abe \emph{et al.}, Phys. Rev. Lett.
  \textbf{88}, 181803 (2002);
  Belle Collaboration, M.Z.~Wang \emph{et al.}, Phys. Rev. Lett.
  \textbf{90}, 201802 (2003);
  Belle Collaboration, M.Z.~Wang \emph{et al.}, Phys. Rev. Lett.
  \textbf{92},131801 (2003);
  Belle Collaboration, Y.J.~Lee \emph{et al.}, Phys. Rev. Lett.
  \textbf{93}, 211801 (2004).
\bibitem{NNG}  Belle Collaboration, Y.J.~Lee \emph{et al.}, Phys. Rev. Lett. 
  \textbf{95}, 061802 (2005).
\bibitem{CC} Inclusion of the charge conjugate state is implied throughout 
             this paper. 
\bibitem{belle} Belle Collaboration, A.Abashian \emph{et al.}, Nucl. Instr.
and Meth. A \textbf{479}, 117 (2002).
\bibitem{kekb}S. Kurokawa and E. Kikutani, Nucl. Instr. and Meth. A 
  \textbf{499}, 1 (2003)
\bibitem{PRD03} Belle Collaboration, K. Abe \emph{et al.}, Phys. Rev.
 \textbf{D67}, 032003 (2003).
\bibitem{FW}G.C. Fox and S. Wolfram, Phys. Rev. Lett. 
 \textbf{41}, 1581 (1978).
\bibitem{pdg}S. Eidelman \emph{et al.} (Particle Data Group),
  Phys. Lett. B \textbf{592}, 1 (2004).
\bibitem{dmb}The benefit of using of $\Dmb$ instead of the energy difference is described in Ref. \cite{zsl}.
\bibitem{ARGUS}  ARGUS Collaboration, H. Albrecht \emph{et al.}, Phys. Lett. B
  \textbf{241}, 278 (1990).
\bibitem{FC}G.J. Feldman and R.D. Cousins, Phys. Rev. D \textbf{57}, 3873 (1998).
\bibitem{Conrad}J. Conrad \emph{et al.}, Phys. Rev. D \textbf{67}, 012002 (2003).
\end{thebibliography}
\end{document}